\documentclass[12pt,preprint]{aastex}
\usepackage{epsf}

\shorttitle{Cosmological Shock Waves}
\shortauthors{Ryu {\it et al.~}}
\def\etal{{\it et al.}}
\def\ie{{\it i.e.,~}}

\begin{document}
\title{Cosmological Shock Waves and Their Role in the Large Scale Structure
of the Universe}

\author{Dongsu Ryu\altaffilmark{1,3},
        Hyesung Kang\altaffilmark{2,3},
        Eric Hallman\altaffilmark{3},
    and T. W. Jones\altaffilmark{3}}

\altaffiltext{1}
{Department of Astronomy \& Space Science, Chungnam National University,
Daejeon 305-764, Korea:\\ ryu@canopus.chungnam.ac.kr}
\altaffiltext{2}
{Department of Earth Sciences, Pusan National University, Pusan 609-735,
Korea:\\ kang@uju.es.pusan.ac.kr} 
\altaffiltext{3}
{Department of Astronomy, University of Minnesota, Minneapolis, MN 55455:\\
hallman@msi.umn.edu, twj@msi.umn.edu}

\begin{abstract}
We study the properties of cosmological shock waves identified in
high-resolution, N-body/hydro\-dynamic simulations of a $\Lambda$CDM
universe and their role on thermalization of gas and acceleration of
nonthermal, cosmic ray (CR) particles. External shocks form around sheets,
filaments and knots of mass distribution when the gas in void regions
accretes onto them. Within those nonlinear structures, internal shocks
are produced by infall of previously shocked gas to filaments and knots,
and during subclump mergers, as well as by chaotic flow motions.
Due to the low temperature of the accreting gas, the Mach number
of external shocks is high, extending up to $M\sim 100$ or higher.
In contrast, internal shocks have mostly low Mach numbers. For all
shocks of $M\ge1.5$ the mean distance between shock surfaces over
the entire computed volume is $\sim4 h^{-1}$ Mpc at present, or
$\sim 1 h^{-1}$ Mpc for internal shocks within nonlinear structures. 
Identified external shocks are more extensive, with their surface area
$\sim2$ times larger than that of identified internal shocks at present.
However, especially because of higher preshock densities, but also due
to higher shock speeds, internal shocks dissipate more energy. Hence,
the internal shocks are mainly responsible for gas thermalization as
well as CR acceleration. In fact, internal shocks with $2 \la M \la 4$
contribute $\sim 1/2$ of the total dissipation. Using a nonlinear
diffusive shock acceleration model for CR protons, we estimate the ratio
of CR energy to gas thermal energy dissipated at cosmological shock
waves to be $\sim1/2$ through the history of the universe. Our result
supports scenarios in which the intracluster medium contains energetically
significant populations of CRs.
\end{abstract}


\section{Introduction}

According to cosmological N-body/hydrodynamic simulations, intergalactic
shock waves develop as a consequence of the large scale structure
formation of the universe. Infall of baryonic gas toward sheets, filaments
and knots, as well as supersonic flows associated with hierarchical
clustering, induce shocks \citep[see, e.g.,][]{qis98,mrk00,mini02,gb03}.
Those cosmological shock waves, like most astrophysical shocks, are
``collisionless'' features mediated by collective, electromagnetic
viscosities. Such viscosities rely on irregular magnetic field components,
\ie MHD wave turbulence that is self-excited by the streaming suprathermal
particles produced during shock formation
\citep[see, e.g.,][]{went74,keh85,quest88}.

The existence and character of these shocks is important for several
reasons. Through dissipation the cosmological shock waves convert part of
the gravitational energy associated with structure formation into heat and,
thus, govern the gas thermal history in the universe. Thermal energy is
then radiated away, manifesting the large scale structure as well
as the dynamics that created it \citep[see, e.g.,][]{co99,dcob01}.
At the same time, due to incomplete plasma thermalization at
collisionless shocks, a sizeable portion of the shock energy can be
converted into cosmic ray (CR) energy (mostly ionic) via diffusive shock
acceleration (DSA) \citep[for reviews, see, e.g.,][]{be87,md01}. This
nonthermal population represents a small fraction of the ion flux
through a collisionless shock that leaks back upstream to become subject
to DSA; that is, to be ``injected'' into the nonthermal population.
Numerous nonlinear studies of DSA have shown that substantial fractions
of the energy flux through strong shocks can be captured by the nonthermal
populations \citep[e.g.,][]{ber95,ell96,mal99,kjg02}. Extensive nonlinear
simulations by some of us incorporating a plasma-physics-based ``thermal
leakage'' injection model into combined gas dynamic/CR diffusion-convection
simulations found that strong shocks transfer up to $\sim 1/2$ of
the initial shock kinetic energy to CRs by this process \citep{kjg02,kj02}.

There is clear evidence that one or more processes energize significant
nonthermal particle populations in and around cosmic structures.
A number of clusters have been found with diffuse synchrotron radio halos
or/and radio relic sources, indicating the existence of relativistic
electron populations in intracluster medium (ICM)
\citep[see, e.g.,][]{gf00,fere03}. In addition, some clusters have been
reported to possess excess EUV and/or hard X-ray radiation compared to
that expected from the hot, thermal X-ray emitting ICM, most likely
produced by inverse Compton scattering of cosmic microwave background
radiation (CMBR) photons by CR electrons
\citep[see, e.g.,][]{lmbl96,fdfg99}. Also it has been suggested that
a fraction of the diffuse $\gamma$-ray background radiation could
originate from the same process \citep{lw00,mini02,sm02}. If some of
those CR electrons have been energized at cosmological shock waves,
the same process should have produced greater CR proton populations.
Hence, although CR protons in the ICM have yet to be confirmed by the
observation of $\gamma$-ray photons produced by inelastic collisions
between CR and thermal-gas protons \citep[see, e.g.,][]{mrkj01,rpsm03},
there may very well exist CR proton populations there whose pressure
is comparable to the gas thermal pressure \citep{ebkw77,cb98,liab99}.

The properties of cosmological shock waves in the large scale structure
of the universe were analyzed quantitatively by \citet{mrk00} using
numerical simulations with $270^3$ grid zones in a cubic comoving region
of size $85h^{-1}$ Mpc for a SCDM model and a $\Lambda$CDM model. They
identified accretion shocks, merger shocks and internal flow shocks which
were formed by infall and hierarchical clustering, and showed that the
topology of these shocks is very complex. They found merger and internal
flow shocks distributed over Mach numbers from 3 to 10 with a peak at
$M \sim 5$ and accretion shock Mach numbers ranging between 10 and a few
$\times 10^3$. In a recent study based on a higher resolution simulation
with $512^3$ grid zones in a cubic comoving region of size $50h^{-1}$
Mpc for a $\Lambda$CDM model, \citet{mini02} showed that most dissipation
involves shocks of modest strength, with $4 \la M \la 10$ accounting for
$\sim 45\%$ of total shock heating. On the other hand, through
a semi-analytic study \citet{gb03} found a Mach number distribution of
merger-related shocks during large scale structure formation with a peak
at much lower Mach number ($M \la 1.5$).

In this paper, we critically re-examine the properties of cosmological
shock waves with a new set of cosmic structure formation simulations. 
We quantify the characteristics of cosmological shock waves and estimate
the dissipation, gas thermalization and CR acceleration at those shocks.
The capture of shocks in hydrodynamic simulations and the identification
of shocks in such simulation data are affected by numerical details
including resolution. Therefore, to validate our findings we estimate
the errors in our measured quantities through consistency checks and
resolution convergence tests. This study adds valuable insights into
the thermal history and nature of gas in the universe, as well as
nonthermal activities in the ICM. In \S 2, simulations are detailed
along with shock identification. The main results of shock
characteristics and shock dissipation are described in \S 3
and \S 4, respectively, followed by a summary in \S 5.

\section{Numerics}

\subsection{Simulations}

The cold dark matter cosmology with a cosmological constant ($\Lambda$CDM)
was employed with the following parameters: $\Omega_{BM}=0.043$,
$\Omega_{DM}=0.227$, and $\Omega_{\Lambda}=0.73$,
$h \equiv H_0$/(100 km/s/Mpc) = 0.7, and $\sigma_8 = 0.8$. The above
values are consistent with those fitted with the recent WMAP data
\citep[see, e.g.,][]{bhhj03}. A cubic region of comoving size $100h^{-1}$
Mpc was simulated inside a computational box with $1024^3$, $512^3$,
$256^3$, $128^3$ and $64^3$ grid zones for gas and gravity and $512^3$,
$256^3$, $128^3$, $64^3$ and $32^3$ particles for dark matter. It allows
a uniform spatial resolution of $\Delta l = 97.7h^{-1}$ kpc $-$
$1.57 h^{-1}$ Mpc. The simulations were performed using a PM/Eulerian
hydrodynamic cosmology code. The code is described in \citet{rokc93}.
But the version of the code used includes several updates. For instance,
it now adopts the MC (monotonized central difference) limiter, instead of
the original minmod limiter. The update to the MC limiter was intended
to enhance the density resolution and to capture shocks more sharply
\citep[see, e.g.,][for details]{leve97}.

We did not include in our simulations several physical processes 
such as radiative cooling, galaxy/star formation, feedback from
galaxies and stars, that can play significant roles in determining
conditions within cluster cores, nor reionization of the intergalactic
medium that effectively sets a temperature floor to the IGM. 
Our primary goal is to study cosmological shocks which are mostly
outside cluster core regions. We established a temperature floor
as a part of our analysis procedure. The conclusions drawn in this study,
hence, should not be significantly weakened by the exclusion of those
additional physical processes.

\subsection{Shock Identification}

While shocks are automatically detected during the simulations by
the Riemann solver within the hydrodynamics routine, there are
additional steps necessary to identify and characterize shocks
for analysis. We have done this as a post-processing step using 
the simulation data at selected epochs. Ideally,
explicitly three-dimensional flow motions should be considered in
identifying shocks in simulation data. However, to simplify the analysis
we used a one-dimensional procedure applied successively in all three 
primary directions. A zone was tagged as a ``shock zone'' currently
experiencing shock dissipation whenever these three criteria are met:
1) $\Delta T \cdot \Delta s > 0$, \ie the gradients of gas temperature
and entropy have the same sign, 
2) $\nabla \cdot {\vec v} < 0,$ \ie the local flow is converging (where 
$\nabla \cdot {\vec v}$ is the divergence of three-dimensional velocity
field), 
3) $|\Delta \log T| \ge 0.11$, where in each case we define
central differences according to the scheme, $\Delta Q = Q_{i+1} - Q_{i-1}$.
The third condition corresponds to the temperature jump of a Mach 1.3 shock.
Typically, a shock is represented by a jump spread over $2-3$ tagged
zones. Hence, we identified a ``shock center'' within the numerical shock
where $\nabla \cdot {\vec v}$ is minimum and labeled this center as part
of a shock surface.

The Mach number of shock centers, $M$, was calculated from the temperature
jump across shocks, which is given by
\begin{equation} 
{T_2 \over T_1} = { {(5M^2-1) (M^2+3)} \over {16M^2} },
\end{equation} 
where $T_2$ and $T_1$ are the postshock and preshock temperatures.
Shock centers identified in multiple directional passes were labeled by
the maximum $M$. We followed only those portions of shock surfaces with
$M\ge1.5$ to avoid confusion from complex flow patterns and shock
surface topologies associated with very weak shocks. In the actual
simulations, the minimum gas temperature was set as the temperature of CMBR,
that is, $ T_{\rm min} = 2.7 (1+z)$, since photoionization and heating
as well as radiative cooling were ignored. However, considering that
significant reionization would have taken place by $z\sim15$
\citep[see, e.g.,][]{hh03}, in post-processing we set the minimum gas
temperature at $T_{\rm min}=10^4$ K. The shock properties described
below include that consistency adjustment.

Figure 1 shows a two-dimensional slice from the $1024^3$ simulation
isolating a typical group with X-ray emission weighted temperature,
$T_x\approx1.3$ keV, and X-ray luminosity, $L_x\approx 4.2\times 10^{43} h$
erg s$^{-1}$. The figure shows the locations of identified shocks along
with the X-ray emissivity, gas temperature, density, and velocity field
distributions. Although the X-ray emissivity distribution looks relatively
smooth and round, there are complex accreting flows around the group
including three sheets (dotted line contours in the temperature contour
map) and one filament (thin solid line contours). A complex topology
of shock surfaces surrounding the group can be seen in the temperature
contour map and through the locations of identified shocks. The group
as well as the associated sheets and filament are bounded by shocks,
but there are several additional shock structures within that represent
a variety of converging flow patterns. To illustrate the limitations of
conventional spherical accretion concepts, we note that this group
provides an example of structures where shocks can form in the core
by organized infall flows accreting from filaments and sheets that
penetrate deep into the center of potential wells. The Mach number
of the surface along the portion of the accretion shock centered at
$(x,y)=(3.5,4.0)h^{-1}$ Mpc ranges over $2.4-4$ with a mean value of
3.2, while that along the portion centered at $(x,y)=(5.5,4.0) h^{-1}$
Mpc ranges $5.3-8.3$ with a mean value of 6.5. Although quite strong
shocks of Mach number of a few, including these, were often found
within $0.5-1h^{-1}$ Mpc from the center of clusters and groups,
most shocks identified in our simulations are located
outside the cores of clusters and groups (see the next section and Figure 5).

In order to illustrate how our shock identification scheme works, we
plot in Figure 2 the flow structure along the horizontal path drawn 
just below $y = 4 h^{-1}$ Mpc in the shock location plot of Figure 1.
As indicated in the upper right panel of the figure, two outer shocks
with high Mach numbers and three inner shocks with lower Mach numbers
were identified along the path. The existence of these five shocks is
also clearly evident in the velocity field and temperature contours
in Figure 1. Extensive tests showed that our scheme identifies shock
surfaces reliably with a typical error of a few percent in Mach number.

\section{Properties of Cosmological Shock Waves}

In previous studies by \citet{mrk00} and by other authors,
cosmological shock waves were often organized into three categories;
\ie accretion shocks, merger shocks, and internal flow shocks.
However, based on close examination of shock locations
and the properties of the shocks and their associated flows
in our simulations, we suggest instead that it is very
informative to classify cosmological shocks into two broad populations
that can be conveniently labeled as {\it external} and {\it internal}
shocks, depending on whether or not the associated preshock gas was
previously shocked (that is, whether $T_1 \le 10^4$ or $> 10^4$ for the
preshock temperature in practice). This binary classification
facilitates the understanding of their role in energy dissipation
(see the next section). {\it External shocks} surround sheets,
filaments and knots, forming  when never-shocked, low density,
void gas accretes onto those nonlinear structures. Subsequent,
{\it internal shocks} are distributed within the regions bounded
by external shocks. They are produced by  flow motions accompanying
hierarchical structure formation inside the bounding shocks.
For more refined questions internal shocks can be further divided
into three types: 1) accretion shocks produced by infall from
sheets to filaments and knots and from filaments to knots, 2) merger
shocks formed during subclump mergers, and 3) flow shocks induced by
chaotic supersonic motions inside the nonlinear structures. 

Figure 3 represents a two dimensional slice of a $(25 h^{-1}$ Mpc)$^3$
volume extracted from the $1024^3$ grid zone simulation. It shows
the distributions of external and internal shocks in a cluster with
X-ray emission-weighted temperature, $T_x\approx 3.3$ keV, and X-ray
luminosity, $L_x\approx 1.4\times10^{45}h$ erg s$^{-1}$, along with
the gas density distribution and the velocity field of the inner
parts of the slice. External shocks here define an entire
``cluster complex'' which has dimensions of about
$(10\times 10 \times 20)(h^{-1} {\rm Mpc})^3$. Numerous internal shocks
were identified inside the complex. The two-dimensional velocity field
demonstrates that infalls from several associated filaments and sheets
form accretion shocks and also induce chaotic flow motions in
the medium around the cluster. Figure 4 shows a three-dimensional view
of shock surfaces around the same cluster complex. Surfaces of external
shocks with high Mach numbers, represented with yellow, encompass
the complex and filaments associated with it. Several sheets with
lower Mach numbers, which intersect at the complex, are also clearly
visible. These figures show clearly that the canonical spherical
external accretion shock model around a cluster would be far too
simple to apply to this sample cluster. Note, in particular, that
a ``spherical'' cluster of $T_x = 3.3$ keV has the first caustic
at $R_c \approx 3.5 h^{-1}$ Mpc from the center in our
$\Lambda$CDM model \citep{rk97}, around which the external shock
would be located. That is well inside the external, accretion shock in
this simulated cluster complex.

We will now estimate some quantitative measures of shock frequency. To
start we computed the surface area of identified shocks per logarithmic
Mach number interval, $dS(M,z)/d\log M$, normalized by the volume of
the simulation box. This provides an effective inverse comoving mean
distance between shock surfaces. In this accounting each shock center
contributes a surface $1.19 (\Delta l)^2$, which is the mean projected
area within a three-dimensional zone for random shock normal orientations. 
The upper two panels of Figure 5 show $dS(M,z)/d\log M$ for external and
internal shocks at several epochs in the simulation with $1024^3$ grid
zones. Table 1 lists the integrated mean shock separation, $1/S(z)$,
along with the mean values of the shock Mach number, the shock speed,
the preshock sound speed and the preshock gas density.

Several important points are apparent.
1) The two populations of shocks have distinctive distributions of shock
surfaces, $dS(M,z)/d\log M$, implying they are induced by flows of different
characteristics. The area distribution of external shock surfaces peaks
at $M \approx 3-5$, extending up to $M\sim 100$ or higher. In contrast,
the comoving area of internal shocks increases to the weakest shocks we
identified in our analysis (\ie $M=1.5$).
2) While the comoving area of external shock surfaces has not changed
much since $z\la 2$, that of internal shocks has increased significantly.
The former behavior reflects the fact that the ``bounding'' shapes
associated with external shocks have evolved towards simpler shapes to
almost balance the increasing comoving volumes enclosed. On the other
hand both the volume enclosed and the complexity of internal shocks has
increased, so that these shocks now include more area. At the present
epoch the total area of external shock surfaces is $\sim 2$ times of
that of identified internal shocks.
3) Although the mean Mach number is higher for external shocks, the
mean shock speed is actually larger for internal shock. In addition,
the preshock gas density is significantly higher for internal shocks.
These factors enhance their dynamical importance (see the next section).
4) We found that most clusters and groups with $ T_x \ga 0.1$ keV have
shocks within $0.5h^{-1}$ Mpc from the centers at present. The area
distribution of these ``cluster shocks'', shown in the upper right
panel of Figure 5, fits best to
\begin{equation} 
{dS(M,z) \over d\log M} \propto \exp \left(-\sqrt{M \over M_{ch}} \right)
\end{equation} 
with $M_{ch}\approx1$ in the range of $M \la 10$. Their mean Mach number
is $\sim 4$. The cluster shocks, however, actually account for only a very
small fraction of identified shock surfaces. We emphasize that the
statistics for the cluster shocks would have been affected by the finite
resolution, $\Delta l = 97.7h^{-1}$ kpc, as well as by the exclusion of
physical processes like radiative cooling and feedback from galaxies
and stars that influence conditions inside cluster cores. Still, it is
significant that, compared with the distribution of binary merger shocks
studied in \citet{gb03}, shocks with higher Mach numbers are more common
in the environments of clusters and groups in our simulation. This
difference is because our cluster shock population includes accretion
shocks created by infall from filaments and sheets to knots and flow
shocks generated by chaotic supersonic motions, as well as those
induced by hierarchical merging. Especially, internal accretion shocks
are strong with a typical Mach number of several (see \S 2.2).

The lower two panels of Figure 5 show $dS(M,0)/d\log M$ for external and
internal shocks at $z=0$ in the simulations with different numerical
resolutions of $1024^3 - 128^3$ grid zones. Table 2 lists the integrated
surface area, $S(0)$, along with the mass of the gas that went through
shocks at least once, ${\cal M}_{sg}(0)$, and the total gas thermal energy
inside the computational box, ${\cal E}_{th}(0)$, at the present epoch in the
simulations with $1024^3 - 64^3$ grid zones. 
The shocked gas mass, ${\cal M}_{sg}$ was estimated by
summing the mass of gas with $T > 10^4$ K. We expect that finite
resolution would affect the statistics of shocks in two different ways:
1) some shocks, especially weak ones with low Mach numbers, may not have
been ``captured'' in the hydrodynamic part of simulations, and 2) they
may not have been ``identified'' by our post-processing shock
identification scheme, especially in the regions with complex
three-dimensional flow structures. The former causes quantities like
${\cal M}_{sg}$ and ${\cal E}_{th}$ to be underestimated. On the other
hand, the latter reduces the estimated value of $S(z)$ and so the
estimation of shock dissipation (see the next section). 
We see that ${\cal M}_{sg}$ and ${\cal E}_{th}$
converge rather quickly and their converged values should be within
$\sim 10\%$ or so of those from the simulation with $1024^3$. This
indicates that the ability of our code to capture shocks is not the
major limitation. The convergence of the mean shock separation,
$1/S$ and $S_{ext}/S_{int}$, is not as obvious from the table. 
The difficulties in reaching final convergence in these quantities
is evident in Figure 5. Detection of strong shocks with $M\ga10$,
especially strong external shocks, is relatively robust in the
simulation data with $256^3$ or more grid zones. However, the area
spanned by weak shocks continues to increase with resolution,
reflecting the increased flow complexity captured inside nonlinear
structures.  The quantities of $1/S$ and $S_{ext}/S_{int}$ are, however, also
plotted in Figure 8, where it is evident that convergence is
underway between the two highest resolution simulations. It is likely
that $1/S \sim 4 h^{-1}$ Mpc, and $S_{ext}/S_{int} \sim 2$. It is,
then, interesting to note that since the volume of the nonlinear
structures where internal shocks are found is $\sim 1/10$ of the entire
computed volume, the mean distance between internal shock surfaces is
$\sim 1h^{-1}$ Mpc within the nonlinear structures, which is comparable
to the scale of the nonlinear structure involved. This conclusion,
of course, includes the caveats that only shocks with $M \ge 1.5$
have been counted and that we have omitted from the simulations
physics likely to influence shock formation inside cluster cores.
Accepting the above limitations, we argue in the next section from
estimations of the gas thermal energy dissipated at shocks
and the mass passed through external shocks that the identification
of dynamically important shocks should be complete within an error of
order $10\%$ in the simulation data with $1024^3$ grid zones.

%While structure
%shock development is certainly not fully converged in our simulations,
%convergence trends are well enough established for several key measures
%that they should be reliable within order $10$\% or so. 

\section{Energy Dissipation at Shocks}

As the first step to quantitative estimates of the gas thermal energy
and CR energy from dissipation at cosmological shock waves, we defined
the following fluxes of mass and energies at each ``shock center'':
1) the gas mass flux incident on shocks, $f_{ms}$ ($= \rho_1 v_{sh}$);
2) the incident kinetic energy flux, $f_{\phi}$ ($= (1/2) \rho_1 v_{sh}^3$);
3) the thermal energy flux generated at shocks, $f_{th}$ (defined below);
and 4) the CR energy extracted, \ie ``nonthermal dissipation'', at shocks,
$f_{CR}$ (defined below). Here subscripts 1 and 2 stand for preshock
and postshock conditions, respectively.

We define thermal energy flux generated at shocks as 
\begin{equation} 
f_{th} = \left[e_{th,2} - e_{th,1} \left({\rho_2\over\rho_1}
\right)^{\gamma}\right]u_2.
\end{equation} 
We point out that the second term inside the brackets subtracts the effect
of adiabatic compression occurred at a shock, not simply the thermal
energy flux entering the shock; namely, $e_{th,1} u_1$. The ratio
$f_{th}/f_{\phi} \equiv \delta (M)$ then defines the efficiency of
shock thermalization, which is a function of Mach number only,
and can be determined from the Rankine-Hugoniot jump conditions.
The left panel of Figure 6 shows $\delta (M)$. As expected,
$\delta (M)$ increases with Mach number, asymptotically approaching
$\delta (M) \rightarrow 0.56$ for $M \ga 10$.

We express the CR energy extraction rate as $f_{CR} = \eta(M) f_{\phi}$,
where $\eta(M)$ measures the efficiency of diffusive shock acceleration
for a given Mach number. To estimate this efficiency we have used results
of numerical simulations of the nonlinear evolution of CR modified
quasi-parallel plane shocks \citep[e.g.,][]{kjg02,kj02}. The simulations
utilize a plasma-physics-based model for the leakage of thermal ions into
the CR population, which depends on a single parameter measuring
the enhancement of nonlinear scattering waves across the shock
transition. That parameter is reasonably well limited by the theory and
by plasma simulations \citep[see, e.g.,][]{mal98}, and, in the
DSA simulation we used, typically limits the fraction of protons injected
into the CR population at shocks to be of order $\sim 10^{-3}$.
The CR acceleration efficiency characteristics of our DSA model are 
broadly consistent with other widely used theoretical and numerical
studies of nonlinear CR shocks \citep[e.g.,][]{mal99,ber95,ell96}.
In order to apply properties of our DSA model to cosmological shock
waves, we calculated proton acceleration and accompanying CR-modified
flow evolution for shocks with $v_{sh}=1500-3000$ km s$^{-1}$ propagating
into media of $T_1 = 10^4 - 10^8$ K, assuming Bohm-type diffusion for
the CRs \citep[for details see][]{kjg02}. As the shock structures evolved
we determined the ratio of the total CR energy extraction during
the evolution of a shock to the kinetic energy passed through the shock
according to
\begin{equation}
\Phi_{CR}(t) \equiv {\int_x E_{CR}(x,t) dx~{\bigg/}~
\left[ {1\over2}\rho_1 v_{sh}^3 t \right]}.
\end{equation}
The values of $\Phi_{CR}(t)$ quickly reached approximate ``time-asymptotic''
values, after which the shock structures evolved approximately
``self-similarly.'' These asymptotic values of $\Phi_{CR}(t)$ were taken as
our estimates for $\eta (M)$. 

The left panel of Figure 6 shows the resulting CR acceleration efficiency,
$\eta (M)$, along with the gas thermalization efficiency $\delta (M)$.
Both $\delta (M)$ and $\eta (M)$ increase with Mach number. In strong,
high Mach number shocks, $\eta (M)$ approaches the asymptotic value of
$\eta (M) \rightarrow \sim 0.53$. We see that $\delta (M)$ is somewhat
larger than $\eta (M)$ over all Mach numbers. On the other hand, we note
that $\delta(M)$ was computed for purely gasdynamical shocks; that is,
without accounting for energy removal into CRs and enhanced adiabatic
compression within the CR shock precursor. If nonlinear dynamical feedback
of CRs were included self-consistently in estimating $\delta (M)$,
the resulting $\delta (M)$ would be somewhat smaller than that shown
in Figure 6. 

From the above gas mass and kinetic energy fluxes at each shock center,
we calculated the associated fluxes through surfaces of shocks with Mach
number between $\log M$ and $\log M + d (\log M)$ at different redshifts;
\ie $dF_{ms}(M,z)$ and $dF_{\phi}(M,z)$, respectively. We also calculated
the similarly defined fluxes of gas thermal and CR energies dissipated at 
shocks as $dF_{th}(M,z) =d F_{\phi}(M,z) \times \delta(M)$ and
$dF_{CR}(M,z) =d F_{\phi}(M,z) \times \eta(M)$, respectively. The right
panel of Figure 6 illustrates $dF_{\phi}(M,z)/d\log M$ per unit comoving
volume for external and internal shocks at different redshifts in the
simulation with $1024^3$ grid zones. A noticeable point is that the kinetic
energy flux, and also the gas mass flux through external shocks were
larger in the past. This is because the preshock gas density was larger
in the past. However, the kinetic energy flux through internal shocks has
been more or less constant since $z\sim1.5$, and was smaller before that.

To provide measures of the roles of shocks over time we integrated
from $z=2$ to $z=0$ the gas mass and kinetic energy that passed through
shock surfaces and the gas thermal and CR energies dissipated at shock
surfaces as
\begin{equation}
{{d Y_i(M)} \over {d \log M}} = { 1 \over {\cal N}_i}
\int_{z=2}^{z=0} {{d F_i(M,z)} \over {d \log M}} dt,
\end{equation}
where the subscript $i \equiv ms~,\phi,~ th,~{\rm and}~CR$ stands for
the four fluxes defined above. The quantities were normalized either
to the shocked gas mass, ${\cal N}_{ms} \equiv {\cal M}_{sg}(z=0)$,
for mass or to the total gas thermal energy inside the computational
box, ${\cal N}_i \equiv {\cal E}_{th}(z=0)$, for energies (see Table 2).
We also summed these time-integrated measures to calculate the associated
global shock-processed quantities,
\begin{equation}
Y_i(>M) = \int_{\infty}^M \left[ {{d Y_i(M)} \over {d \log M}} \right]
d \log M.
\end{equation}
$Y_{ms}$ and ${d Y_{ms}(M)} / { d \log M}$ not only measure the total 
mass that passed through shocks in the $z=2$ to $z=0$ interval, but also,
by way of their normalization, measure the mean number of times that gas
has been subjected to shock dissipation. Meanwhile, $Y_{\phi}$ and
its derivative measure the total kinetic energy that has been subject
to shock dissipation. On the other hand, $Y_{th}$ and $Y_{CR}$ compare
the total thermal and CR energies that have resulted from shock dissipation.

Figure 7 shows ${d Y_i(M)} / { d \log M}$ and $Y_i(>M)$ for external
and internal shocks in the simulation with $1024^3$ grid zones.
Again, several important points are apparent.
1) The plots for ${d Y_{ms}(M)} / { d \log M}$ and $Y_{ms}(>M)$ indicate
that more mass has passed through internal shocks than external shocks.
With $Y_{ms}(\ge1.5) \approx 2.2$ for internal shocks, the mass inside the
nonlinear structures of sheets, filaments and knots has been shocked, on
average, twice or so by internal shocks from $z=2$ to 0. But we note that
${d Y_{ms}(M)} / { d \log M}$ for internal shocks increases to $M = 1.5$,
the lowest Mach number we kept, and weak shock identification is not fully
converged yet with this numerical resolution. So the above value of 2.2
should be regarded as the lower limit.
2) On the other hand, since gas enters the nonlinear structures by 
passing through external shocks, $Y_{ms}(\ge1.5)$ for external shocks
(\ie the mass passed through external shocks from $z=2$ to 0) should
match the increase in the shocked gas mass, ${\cal M}_{sg}$, from $z=2$
to 0. In the simulation with $1024^3$ grid zones, we get
$Y_{ms}(\ge1.5) = 0.35$ for external shocks, while $0.42= (0.73-0.42)/0.73$ 
is expected with ${\cal M}_{sg}(z=2) = 0.42{\cal M}_{gas}$ and 
${\cal M}_{sg}(z=0) = 0.73{\cal M}_{gas}$ (where ${\cal M}_{gas}$ is the
total gas mass in the computational box).  Although a fraction
of the hot gas may have been heated above $10^4$ K by adiabatic
compression in void regions or by weak shocks with $M < 1.5$, some of
this discrepancy could be due to the under-counting of external shocks.
So we estimate an error of $\sim 10\%$ or so in the completeness of
the identification of external shocks.
3) The lower two panels of Figure 7 show that internal shocks play a
more important role than external shocks in the dissipating energy associated
with structure formation. Specifically, internal shocks with $2 \la M \la4$
account for $\sim 1/2$ of dissipation. While the thermal energy generation
peaks for shocks in the range $2 \la M \la3$, the CR energy extraction
peaks in the range $3 \la M \la 4$. \citet{mini02} identified shocks with
$4 \la M \la 5$ as the most important in gas thermalization. Our result
indicates a somewhat lower Mach number range, because we found more
weaker shocks in our simulation data. This difference is probably due to
our greater resolution (see, e.g., Figure 5), as well as due to improved
shock capturing in the present code and a more sophisticated shock
identification scheme.
4) The amount of thermal energy generated at shocks from $z=2$
to 0 was computed to be $\sim1.5$ times the current gas thermal energy
in the simulation with $1024^3$ grid zones. Since no other heating or cooling
physics was included, there are two likely contributors to the difference. 
First, we attribute some of the difference to adiabatic expansion of nonlinear
structures after formation. Expansion of $\sim8.5\%$ along each dimension
would be enough to reduce the gas thermal energy by a factor of 1.5.
In addition, some of the discrepancy could be the result of multiple
counting of internal shocks in the regions with complex flow structures.
So again, we put an error of a few $10\%$, at most, in the completeness
of the identification of shocks.
5) With the DSA model we adopted \citep{kjg02, kj02} the ratio of
the CR to gas thermal energies dissipated at cosmological shock waves 
with the Mach number greater than 1.5 is
$Y_{CR}(\ge1.5)/Y_{th}(\ge1.5) \approx 1/2$.
6) There are many shocks with $M \la 2$, but they are not important in
energetics. Hence, although we ignored the shocks with $M < 1.5$ in
this work, that should not significantly impact quantitative energetics
results presented here.

Figure 8 shows $Y_i(\ge1.5)$ through all shock surfaces in the simulations
with different resolution of $1024^3 - 64^3$ grid zones. Note that $Y_i$'s
were normalized with ${\cal M}_{sg}$ and ${\cal E}_{sg}$ for a given resolution
(Table 2). Yet, for instance, $Y_{\phi}$, the integrated kinetic energy
flux through shock surfaces, decreases in the plot by $60\%$ from $1024^3$
to $512^3$ and by $95\%$ from $512^3$ to $256^3$. On the other hand,
${\cal E}_{sg}$, the total gas thermal energy inside the computational box,
decreases only by $8\%$ from $1024^3$ to $512^3$ and $13\%$ from $512^3$
to $256^3$. This reinforces the statements in the previous section that
the error in quantitative assessments of this paper comes mostly from
shock identification in the post-processing analysis rather than shock
capturing in the code. This also shows that our estimation of an error
of order $10\%$ in the data of the $1024^3$ simulation is consistent with
resolution convergence.
 
\section{Summary}

We identified and studied shock waves with Mach number $M\ge1.5$ in a set
of cosmological N-body/hydro\-dynamic simulations for a $\Lambda$CDM
universe in a cubic box of comoving size $100h^{-1}$ Mpc. To facilitate the
analysis of their properties, the cosmological shock waves were classified
as external and internal shocks, depending on whether or not the preshock
gas was previously shocked. External shocks form around outermost surfaces
that encompass nonlinear structures, so they are by nature accretion
shocks that decelerate the previously unshocked intergalactic gas infalling
toward sheets, filaments and knots. Internal shocks are produced within
those nonlinear structures by accretion flows of previously shocked gas
from sheets to filaments and knots and from filaments to knots, by merging
of subclumps, or by chaotic flow motions induced in the course
of hierarchical clustering.

For all shocks of $M\ge1.5$ identified in the simulation of highest
resolution, the mean distance between shock surfaces, the inverse indicator
of shock occurrence, is $\sim 4h^{-1}$ Mpc at present. Further, external
shocks are more extensive, with their surface area $\sim 2$ times larger
than that of identified internal shocks at present. With the volume of
nonlinear structures which is $\sim 1/10$ of the total volume, the mean
distance between internal shock surfaces is $\sim 1h^{-1}$ Mpc within
nonlinear structures at present. Although external shocks typically
have higher Mach numbers, internal shocks have higher shock
speed and higher preshock gas density. As a result, internal shocks are
responsible for $\sim 95 \%$ of gas thermalization and for $\sim 90 \%$
of CR acceleration at shocks, and they process about 6 times more gas mass
through shock surfaces than external shocks do from $z=2$ to $z=0$.
Internal shocks with $2\ \la M \la 4$ are especially important in energy
dissipation, contributing $\sim 1/2$ of the total. By adopting a model of CR
proton acceleration based on nonlinear diffusive shock simulations
\citep{kjg02}, our study predicts that the ratio of the CR to gas thermal
energies dissipated at all cosmological shocks through the history of
the universe could be $\sim1/2$. Due to long CR proton trapping times
and energy loss lifetimes, they should fill the volumes inside filaments
and sheets as well as in clusters and groups. Short electron lifetimes,
however, lead that population to depend on other factors \citep{mjkr01}.
The existence of substantial CR populations could have affected
the evolution and the dynamical status of the large scale structure of
the universe \citep[see, e.g.,][]{ebkw77}.

The examination of results from simulations of different resolutions
showed that the convergence with resolution in shock identification is
slower than in shock capture within the simulation itself. Consistency
checks and resolution convergence analysis lead to error estimates of
order of $10\%$ in our quantitative estimates of the accounting of and
energy dissipation in cosmological shock waves.

Strong, external shocks are energetically less important than moderate
strength internal shocks. However, it is important to keep in mind that
the large curvature radii and long life times of these shocks make them
viable candidates to accelerate CRs to ultra high energies of several
$\times 10^{19}$ eV \citep{norman95,krj96}.

The Mach number distribution and the amount of energy dissipation at
cosmological shocks have significant implications for several cosmological
observations such as radio and $\gamma$-ray emissions as well as X-ray
emission from the ICM and contribution to the cosmic $\gamma$-ray
background from CRs accelerated at these shocks
\citep[see, e.g.,][]{lw00,mrkj01,mjkr01,mini02}. These important issues,
along with the observational manifestation of cosmological shock waves,
will be considered in future studies.  

\acknowledgements
DR and HK were supported by KOSEF through Astrophysical Research Center
for the Structure and Evolution of Cosmos (ARCSEC). EH and TWJ were 
supported by the NSF (AST00-71167), NASA (NAG5-10774) and the University
of Minnesota Supercomputing Institute. Numerical simulations utilized
``The Grand Challenge Program'' of the KISTI Supercomputing Center.
We thank Peter L. Biermann and an anonymous referee for constructive
comments on the manuscript.

\clearpage

\begin{deluxetable}{lcccccccccc}
\tablenum{1}
\tablecaption{Mean flow quantities of external/internal shocks
at several different epochs}
\tablehead{\colhead{$z$} & \colhead{$1/S$}\tablenotemark{a}
& \colhead{$S_{\rm ext}/S_{\rm int}$}
& \colhead{$\langle M \rangle_{\rm ext}$}
& \colhead{$\langle M \rangle_{\rm int}$}
& \colhead{$\langle v_{sh} \rangle_{\rm ext}$}\tablenotemark{a}
& \colhead{$\langle v_{sh} \rangle_{\rm int}$}\tablenotemark{a}
& \colhead{$\langle c_s \rangle_{\rm ext}$}\tablenotemark{a}
& \colhead{$\langle c_s \rangle_{\rm int}$}\tablenotemark{a}
& \colhead{$\langle \rho_{sh} \rangle_{\rm ext}$}\tablenotemark{a}
& \colhead{$\langle \rho_{sh} \rangle_{\rm int}$}\tablenotemark{a}}
\startdata
0   & 4.4 & 2.1 & 8.0 & 3.2 & 123 & 226 & 15.3 & 82 & 1.05 & 6.78 \\
0.2 & 4.4 & 2.3 & 8.1 & 3.3 & 123 & 230 & 15.3 & 83 & 1.12 & 7.15 \\
0.5 & 4.5 & 2.8 & 8.0 & 3.3 & 122 & 231 & 15.3 & 83 & 1.25 & 7.86 \\
1   & 5.0 & 3.7 & 7.5 & 3.4 & 114 & 214 & 15.3 & 76 & 1.48 & 8.87 \\
1.5 & 5.7 & 5.0 & 7.0 & 3.4 & 107 & 196 & 15.3 & 69 & 1.79 & 10.3 \\
2   & 6.8 & 6.6 & 6.5 & 3.4 & 100 & 177 & 15.3 & 62 & 2.14 & 10.9 \\
\enddata
\tablenotetext{a}{Lengths in units of $(1+z)^{-1}h^{-1}{\rm Mpc}$, speeds
in km s$^{-1}$, and density compared to the mean comoving density of gas
$\langle \rho_{gas}\rangle(z)$, respectively.}
\end{deluxetable}

\begin{deluxetable}{rcccc}
\tablenum{2}
\tablecaption{Shock Associated Quantities Measured with Different Resolutions}
\tablehead{\colhead{resolution} & \colhead{$1/S$}\tablenotemark{a}
& \colhead{$S_{\rm ext}/S_{\rm int}$}
 & \colhead{${\cal M}_{sg}$}\tablenotemark{a}
& \colhead{${\cal E}_{th}$}\tablenotemark{a}}
\startdata
$1024^3$ & 4.4 & 2.1 & 0.73 & 1.42 \\
$512^3$  & 7.3 & 3.2 & 0.69 & 1.30 \\
$256^3$  & 14  & 5.4 & 0.61 & 1.13 \\
$128^3$  & 39  & 9.6 & 0.45 & 0.87 \\
$64^3$   & 260 & 28  & 0.19 & 0.40 \\
\enddata
\tablenotetext{a}{Lengths in units of $h^{-1}{\rm Mpc}$, mass compared to
total gas mass inside the computational box ${\cal M}_{gas}$, and energy
in units $10^{64}h^{-1}$ ergs, respectively.}
\end{deluxetable}

\clearpage

\begin{figure}
\vspace{10.cm}
\figcaption{X-ray emissivity contours (upper left panel), velocity field
superimposed on density contours (upper right), gas temperature
contours (lower left) and shock locations (lower right) in a
two-dimensional slice of ($9.76h^{-1}$ Mpc)$^2$ around a group of X-ray
emission weighted temperature $T_x \approx 1.3 $ keV at $z=0$. Contours of
gas density $\rho_{gas}/\langle\rho_{gas}\rangle \ge 1$ are shown.
In the temperature contours, heavy solid lines are used for $\log T> 6.8$,
light solid lines are for $5.5<\log T<6.8$, and dotted lines are for
$4<\log T<5.5$. Several sheets and a filament were identified in
the temperature contours.}
\end{figure}

\clearpage

\begin{figure}
\vspace{-4cm}
\centerline{\epsfxsize=16cm\epsfbox{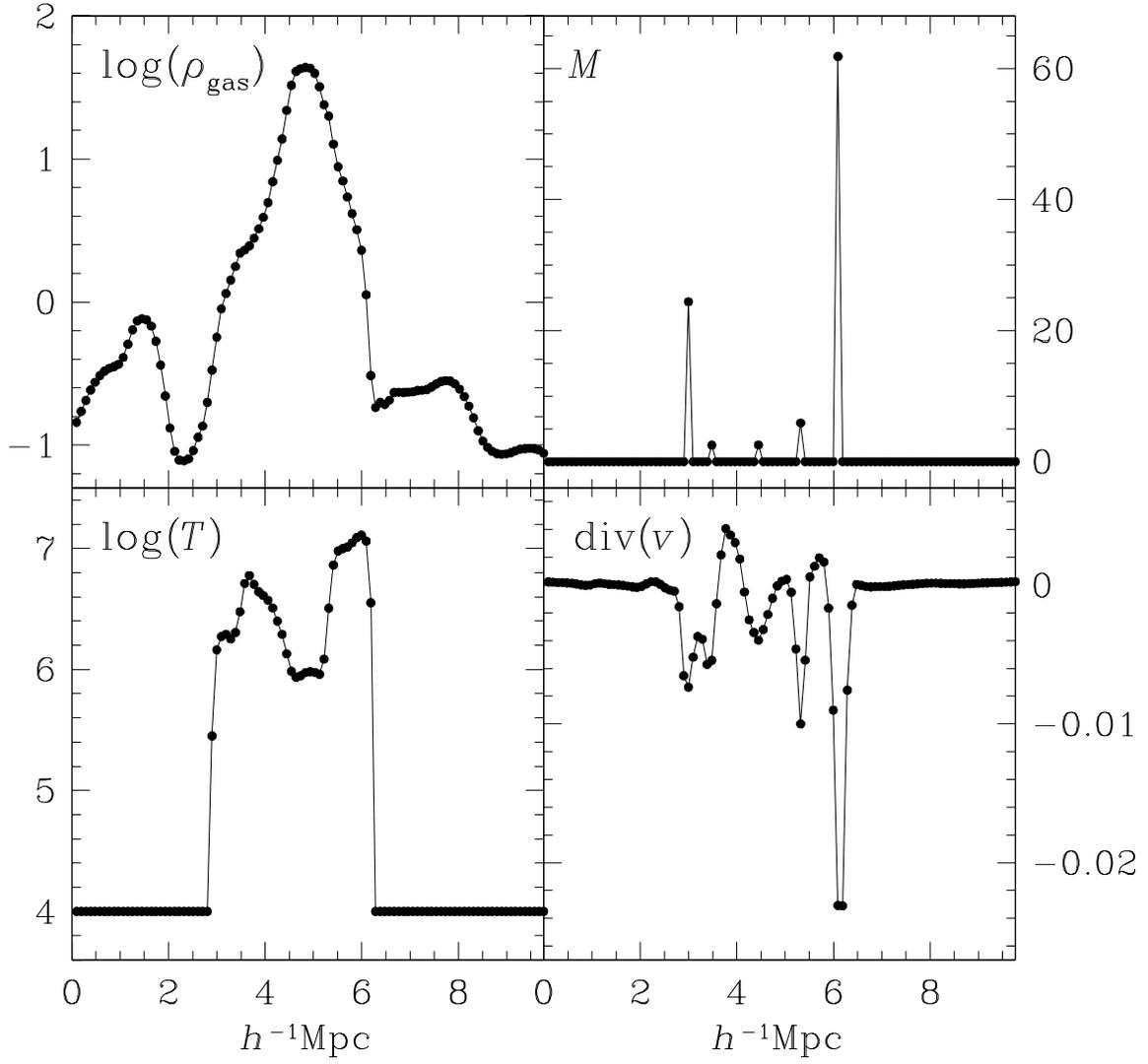}}
\vspace{-2cm}
\figcaption{Flow structure along the line path drawn in Figure 1, showing
gas density, Mach number of identified shocks, gas temperature and
$\nabla \cdot {\vec v}$.}
\end{figure}

\clearpage

\begin{figure}
\vspace{10cm}
\figcaption{Two-dimensional slice of $(25 h^{-1}$Mpc)$^2$ around a
complex including a cluster of X-ray emission weighted temperature
$T_x \approx 3.3$ at $z=0$, showing gas density, internal and external
shock distributions, and velocity field. For clarity, the velocity field
is shown in a zoomed region of ($12.5h^{-1}$ Mpc)$^2$ centered at
the cluster.}
\end{figure}

\clearpage

\begin{figure}
\vspace{10cm}
\figcaption{Three-dimensional shock surfaces in a volume of
$(25 h^{-1}$Mpc)$^3$ around the same complex as in Figure 3. The color
bar shows the values of Mach numbers of shock surfaces.}
\end{figure}

\clearpage

\begin{figure}
\vspace{-4cm}
\centerline{\epsfxsize=16cm\epsfbox{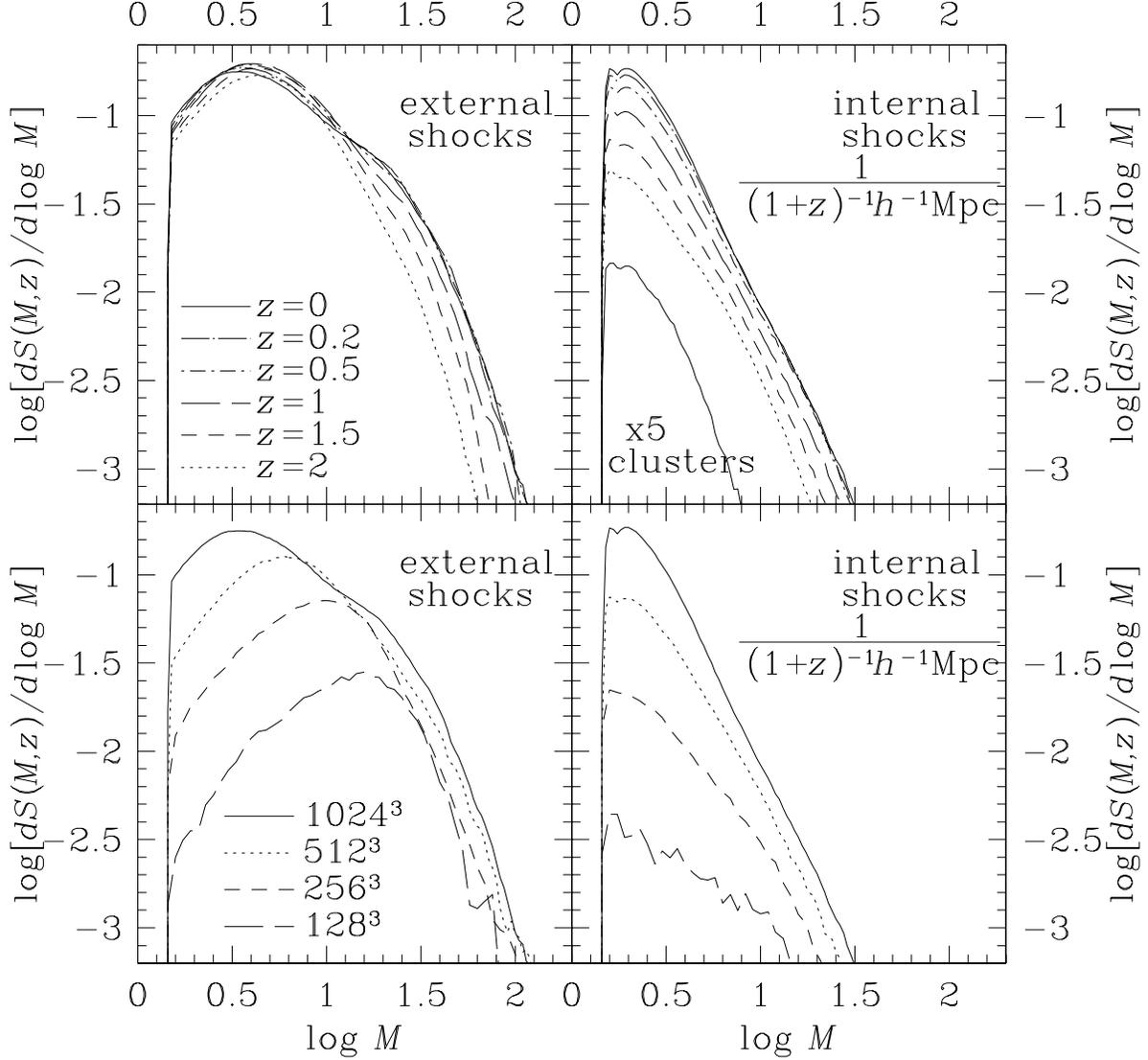}}
\vspace{-2cm}
\figcaption{{\it Top two panels}: Inverse of the mean comoving distance
between shock surfaces with Mach number between
$\log M$ and $\log M + d (\log M)$ at different redshifts, $dS(M,z)$,
for external shocks and internal shocks in the simulation with $1024^3$
grid zones. The curve labeled by ``clusters'' shows the quantity for shocks
inside clusters and groups at present, which was multiplied with 5
for clarity (see the text for details).
{\it Bottom two panels}: The same quantity at $z=0$ in the simulations
with different resolutions of $1024^3 - 128^3$.}
\end{figure}

\clearpage

\begin{figure}
\vspace{-4cm}
\centerline{\epsfxsize=16cm\epsfbox{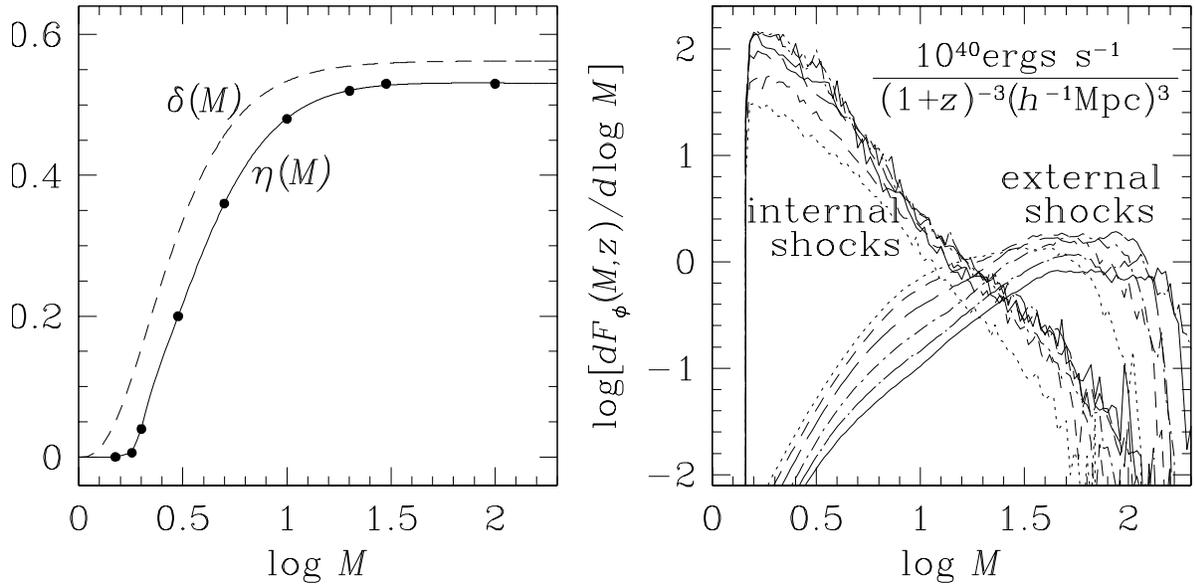}}
\vspace{-2cm}
\figcaption{{\it Left panel}: Gas thermalization efficiency, $\delta(M)$,
and CR acceleration efficiency, $\eta(M)$, at shocks as a function of
Mach number. Dots for $\eta(M)$ are the values estimated from numerical
simulations based on a DSA model and solid line is the fit.
{\it Right panel}: Kinetic energy flux per unit comoving volume through
surfaces of external and internal shocks with Mach number between
$\log M$ and $\log M + d (\log M)$ at different redshifts, $dF_{\phi}(M,z)$,
in the simulation with $1024^3$ grid zones. The line types are same
as those in the upper panels of Figure 5}
\end{figure}

\clearpage

\begin{figure}
\vspace{-4cm}
\centerline{\epsfxsize=16cm\epsfbox{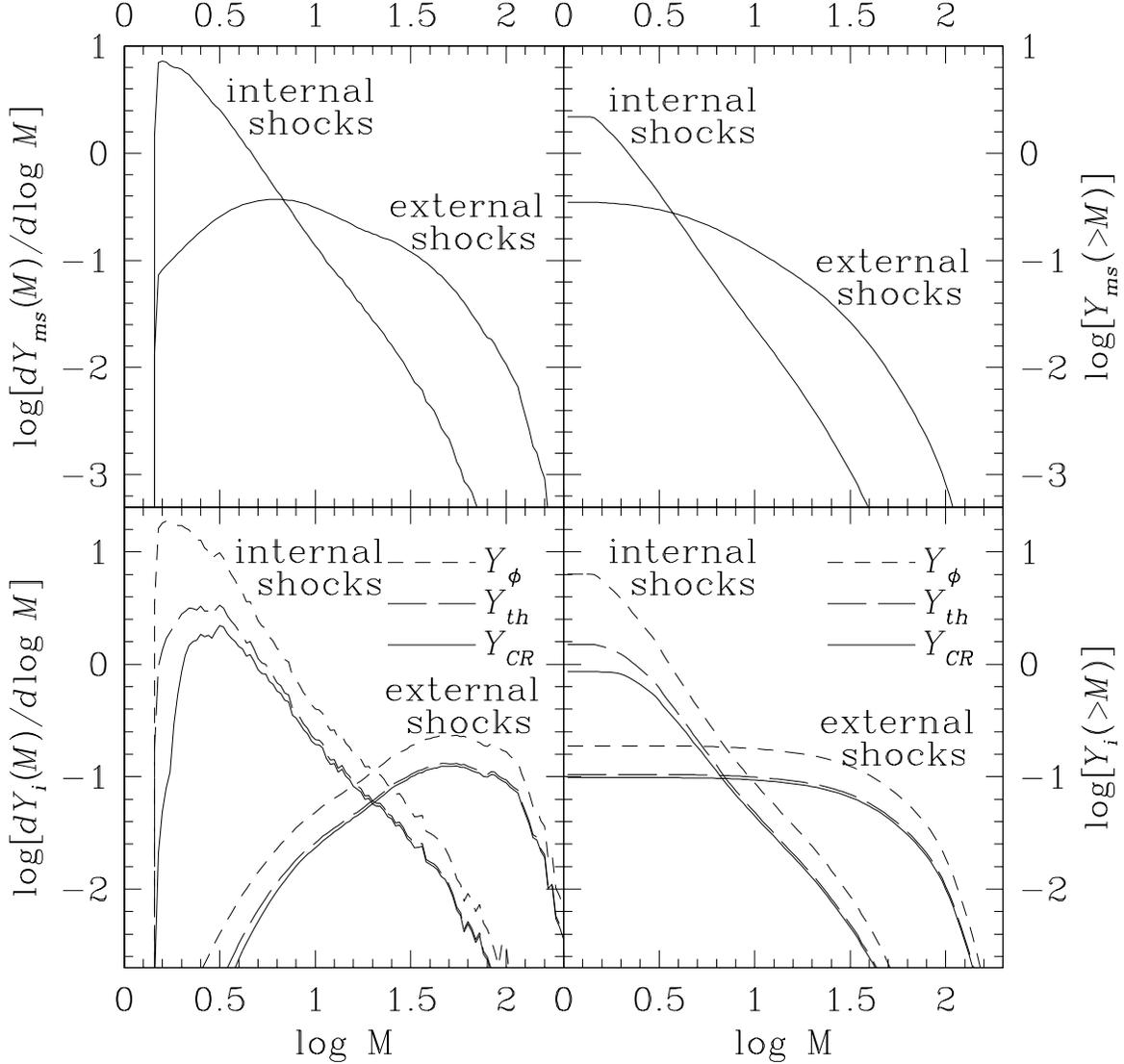}}
\vspace{-2cm}
\figcaption{{\it Upper left panel}: Mass, $dY_{ms}(M)$, processed through
surfaces of external and internal shocks with Mach number between $\log M$
and $\log M + d (\log M)$, from $z=2$ to $z=0$.
{\it Upper Right panel}: Mass, $Y_{mass}(>M)$, processed through surfaces 
of external and internal shocks with Mach number greater than $M$, from
$z=2$ to $z=0$.
{\it Lower left panel}: Kinetic energy, $dY_{\phi}(M)$, thermal energy,
$dY_{th}(M)$, and CR energy, $dY_{CR}(M)$, processed through surfaces
of external and internal shocks with Mach number between $\log M$  and
$\log M + d (\log M)$, from $z=2$ to $z=0$.
{\it Lower Right panel}: Same energies, $Y_{\phi}(>M)$, $Y_{th}(>M)$,
$Y_{CR}(>M)$, processed through surfaces of external and internal
shocks with Mach number greater than $M$, from $z=2$ to $z=0$.
The mass and energies are normalized to the the shocked gas mass,
${\cal M}_{sg}$, and the total gas thermal energy, ${\cal E}_{th}$,
inside simulation box at $z=0$, respectively. All plots are from
the simulation data with $1024^3$ grid zones.}
\end{figure}

\clearpage

\begin{figure}
\vspace{-4cm}
\centerline{\epsfxsize=16cm\epsfbox{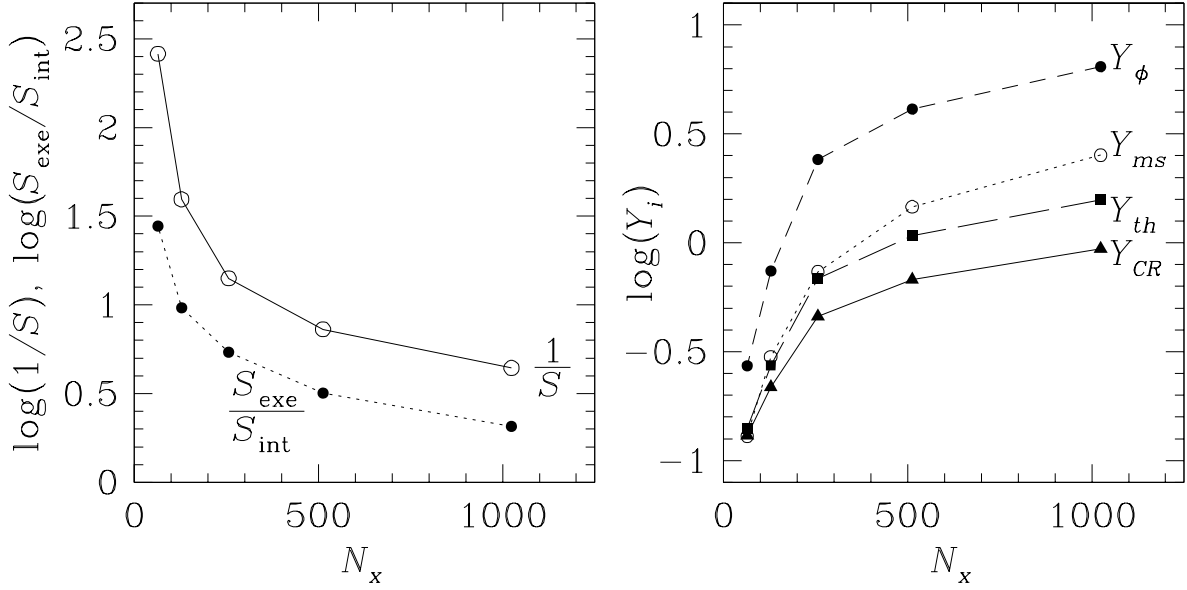}}
\vspace{-2cm}
\figcaption{ {\it Left panel}: Mean shock separation at $z=0$, $1/S$,
and its ratio for two shock populations, $S_{ext}/S_{int}$, for
identified shocks with $M \ge 1.5$ in the simulations of different
resolutions with number of grid points along one-dimension,
$N_x =$ 64, 128, 256, 512, and 1024.
{\it Right panel}: Integrated mass, $Y_{ms}$, and energies, $Y_{\phi}$,
$Y_{th}$, $Y_{CR}$, processed through all identified shocks with
$M \ge 1.5$ in the simulations of different resolutions.
The mass and energies are normalized to the shocked gas mass,
${\cal M}_{sg}$, and the total gas thermal energy, ${\cal E}_{th}$,
of given resolution, inside simulation box at $z=0$, respectively.}
\end{figure}

\end{document}